\documentclass[conference,letterpaper]{IEEEtran}

\usepackage{subeqnarray}
\usepackage{mathrsfs}
\usepackage{amsfonts}
\usepackage{amssymb}
\usepackage{amstext}
\usepackage{graphicx}
\usepackage{indentfirst}
\usepackage{array}
\usepackage{cite}
\usepackage[numbers,sort&compress]{natbib}
\usepackage{enumerate}
\usepackage{stmaryrd}
\usepackage[linesnumbered,ruled,vlined]{algorithm2e}
\usepackage{multirow}
\usepackage{multicol}
\usepackage{verbatim}
\usepackage{stfloats}
\usepackage{color}
\usepackage{bm}
\usepackage{multirow}
\usepackage{multicol}
\usepackage[utf8]{inputenc}
\usepackage[T1]{fontenc}
\usepackage{url}
\usepackage{ifthen}
\usepackage{cite}
\usepackage[cmex10]{amsmath} 

\newtheorem{theorem}{\bf{Theorem}}
\newtheorem{lemma}{\bf{Lemma}}

\newtheorem{definition}{\bf{Definition}}
\newtheorem{remark}{\bf{Remark}}
\newtheorem{example}{\bf{Example}}
\usepackage{fancyhdr}

\begin{document}
\title{A Mathematical Theory of Semantic Communication: Overview}


\author{%
  \IEEEauthorblockN{Kai Niu,  Ping Zhang}
  \IEEEauthorblockA{The State Key Laboratory of Networking and Switching Technology\\
                                 Beijing University of Posts and Telecommunications\\
                                 Beijing, China\\
                                 Email: \{niukai,pzhang\}@bupt.edu.cn}
}


\maketitle
\thispagestyle{fancy}


\begin{abstract}
Semantic communication initiates a new direction for future communication. In this paper, we aim to establish a systematic framework of semantic information theory (SIT). First, we propose a semantic communication model and define the synonymous mapping to indicate the critical relationship between semantic information and syntactic information. Based on this core concept, we introduce the measures of semantic information, such as semantic entropy $H_s(\tilde{U})$, up/down semantic mutual information $I^s(\tilde{X};\tilde{Y})$ $(I_s(\tilde{X};\tilde{Y}))$, semantic capacity $C_s=\max_{p(x)}I^s(\tilde{X};\tilde{Y})$, and semantic rate-distortion function $R_s(D)=\min_{p(\hat{x}|x):\mathbb{E}d_s(\tilde{x},\hat{\tilde{x}})\leq D}I_s(\tilde{X};\hat{\tilde{X}})$. Furthermore, we prove three coding theorems of SIT, that is, the semantic source coding theorem, semantic channel coding theorem, and semantic rate-distortion coding theorem. We find that the limits of information theory are extended by using synonymous mapping, that is, $H_s(\tilde{U})\leq H(U)$, $C_s\geq C$ and $R_s(D)\leq R(D)$. All these works composite the basis of semantic information theory. In summary, the theoretic framework proposed in this paper is a natural extension of classic information theory and may reveal great performance potential for future communication.
\end{abstract}

\section{Introduction}
Classic information theory (CIT), established by C. E. Shannon \cite{Classicpaper_Shannon} in 1948, achieved great success in the information processing and transmission. This theory introduces four critical measures, such as entropy, mutual information, channel capacity, and rate-distortion function to evaluate the performance of information system and proves three famous coding theorems, such as, lossless/lossy source coding theorem and channel coding theorem, to indicate the fundamental limits of data compression and information transmission. Over the past 70 years, information and communication technologies guided by CIT have reached a lot of milestones and approached the theoretical limitation. However, the performance improvement of current communication systems encounters a lot of bottlenecks.

As stated by Weaver in \cite{Semantic_Weaver}, communication involves three level problems:
\begin{enumerate}[]
  \item  ``\textbf{LEVEL A}. How accurately can the symbols of communication be transmitted? (The technical problem.)
  \item  \textbf{LEVEL B}. How precisely do the transmitted symbols convey the desired meaning? (The semantic problem.)
  \item  \textbf{LEVEL C}. How effectively does the received meaning affect conduct in the desired way? (The effectiveness problem.)''
\end{enumerate}
Shannon \cite{Classicpaper_Shannon} pointed that ``semantic aspects of communication are irrelevant to the engineering problem''. Thus, classic information theory only handles LEVEL A (technical) problem of the information.

Many works focused on LEVEL B problem and the semantic communication theory. Carnap and Bar-Hillel \cite{Semantic_Carnap} and Floridi \cite{Semantic_Floridi} considered using propositional logic sentences to express semantic information. Then Bao \emph{et al.} \cite{Semantic_Bao} extended this theoretical framework and derived the semantic source coding and semantic channel coding theorem based on propositional logic probabilities. On the other hand, De Luca \emph{et al.} \cite{Entropy_Luca} \cite{Fuzzy_Luca} regarded semantic information as fuzzy variable and defined fuzzy entropy to measure the uncertainty of semantic information. Then Wu \cite{Wuweiling} extended this work and introduced general entropy and general mutual information based on fuzzy variable. Furthermore, some recent works \cite{RateD_Liu, SideInfo_Guo, Theory_Shao, Theory_Tang} investigated the theoretic property of semantic information. However, designing semantic communication system still lacks of systematic guiding theory.

Recently, as surveyed in \cite{Survey_Gunduz, Semantic_Niu}, semantic communication techniques become a hot topic and a lot of works \cite{Semantic_ZhangPing, Survey_Shi, Survey_Qin, Survey_Xie} investigated the design principles and technical challenges of semantic communication. However, semantic communication research faces a paradox. We can neither precisely answer what is the semantic information nor provide the fundamental limits of the semantic communication system. It is vitally necessary to construct a mathematical theory of semantic communication in order to solve these basic problems.

In \cite{SemInfo_theory}, we found that synonym is the critical feature of semantic information and introduced synonymous mapping to indicate the relationship between the semantic and syntactic information.

As an example of synonym, let's investigate the following two sentences
\begin{enumerate}[]
  \item  ``She appeared \textbf{happy} after receiving the good news''.
  \item  ``She appeared \textbf{joyful} after receiving the good news''.
\end{enumerate}
From the viewpoint of linguistics,  ``happy'' and ``joyful'' are synonym, that is, they have the same meaning. Although these two sentences have different presentations, we can regard them as the same semantic information. Generally, such synonym phenomenon ubiquitously exists in various sources, such as text, speech, image and video. Many different presentations of sentence, waveform, segmentation-object, video frame sequence have the same or similar meaning and compose the synonymous mapping to indicate the same semantic information.

Start from this basic concept, we established a mathematical framework of semantic information theory \cite{SemInfo_theory}. We built the semantic information measures including semantic entropy, up/down semantic mutual information, semantic channel capacity, and semantic rate distortion function. Then we extended the asymptotic equipartition property to the semantic sense and introduced the synonymous typical set to prove three significant coding theorems, that is, semantic source coding theorem, semantic channel coding theorem, and semantic rate distortion coding theorem.
\section{Semantic Communication Model and Synonymous Mapping}
\begin{figure}[htbp]
\setlength{\abovecaptionskip}{0.cm}
\setlength{\belowcaptionskip}{-0.cm}
  \centering{\includegraphics[scale=0.8]{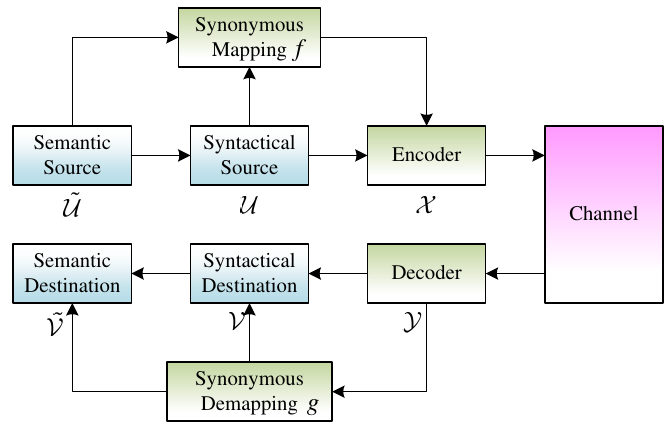}}
  \caption{The block diagram of semantic communication system.}\label{Semantic_communication_system}
\end{figure}

The system model of semantic communication is presented in Fig. \ref{Semantic_communication_system}. The semantic source $\tilde U$ generates a semantic sequence and drives the syntactical source $U$ produces a message. Under the synonymous mapping $f$, the encoder transforms the syntactical message into a codeword and sends to the channel. On the other hand, the decoder recovers the codeword from the received signal and delivers it into the syntactical destination $V$ to reconstruct the message. Then the semantic destination $\tilde V$ reasons the meaning by using such message.

Now we formally introduce the definition of synonymous mapping as follows.
\begin{definition}
Given a syntactic information set $\mathcal{U}=\left\{u_1,\cdots,u_i,\cdots,u_N\right\}$ and the corresponding semantic information set $\tilde {\mathcal{U}}=\left\{\tilde{u}_1,\cdots,\tilde{u}_{i_s},\cdots,\tilde{u}_{\tilde{N}}\right\}$, the synonymous mapping $f:\tilde{\mathcal{U}}\to\mathcal{U}$ is defined as the one-to-many mapping between $\tilde {\mathcal{U}}$ and $\mathcal{U}$. For an arbitrary element $\tilde{u}_i\in \tilde{\mathcal{U}},i_s=1,2,\cdots, \tilde{N}$, we have $f:\tilde{u}_{i_s}\to\mathcal{U}_{i_s}$.
\end{definition}

Generally, the size of set $\tilde {\mathcal{U}}$ is no more than $\mathcal{U}=\left\{u_1,u_2,\cdots,u_N\right\}$, that is, $\tilde{N}\leq N$. Furthermore, under the synonymous mapping $f$, $\mathcal{U}$ is partitioned into a group of synonymous sets $\mathcal{U}_{i_s}=\left\{u_{N_{[1:(i_s-1)]}+1},\cdots,u_{N_{[1:(i_s-1)]}+j},\cdots,u_{N_{[1:(i_s-1)]}+N_{i_s}}\right\}$ where $N_{[i_1:i_m]}$ denotes the integer summation of $N_{i_1}+\cdots+N_{i_j}+\cdots+N_{i_m},\forall N_{i_j}\in \mathbb{N}$ and $\forall i_s\neq j_s,\mathcal{U}_{i_s}\bigcap\mathcal{U}_{j_s}=\varnothing$. Therefore, we have $\left|\mathcal{U}_{i_s}\right|=N_{i_s}$ and $\mathcal{U}=\bigcup_{i_s=1}^{\tilde{N}}\mathcal{U}_{i_s}$.

Essentially, the synonymous mapping $f$ generates an equivalence class partition of the syntactic set. So we can construct the quotient set $\mathcal{U}/f=\left\{\mathcal{U}_{i_s}\right\}$.

\begin{figure}[htbp]
\setlength{\abovecaptionskip}{0.cm}
\setlength{\belowcaptionskip}{-0.cm}
  \centering{\includegraphics[scale=0.7]{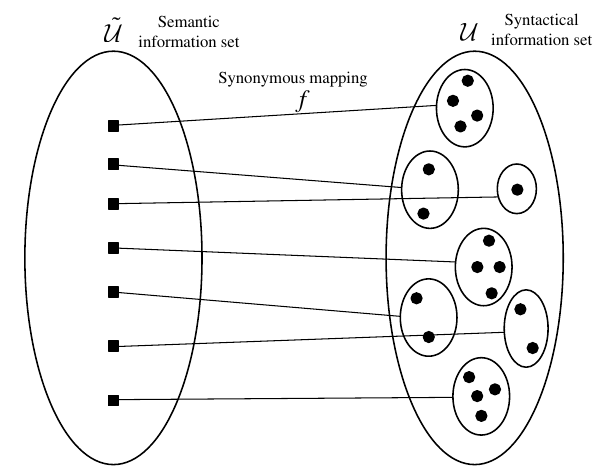}}
  \caption{An example of synonymous mapping between the semantic information set and the syntactic information set.}\label{Synonymous_mapping}
\end{figure}

Figure \ref{Synonymous_mapping} depicts an example of synonymous mapping. Each semantic element can be mapped into an equivalent set of syntactic elements and every set has one or many elements. Further, there is no overlap between any two sets.

\begin{remark}
In fact, the semantic variable $\tilde{U}$ is also a random variable. However, we emphasize that it is implied behind the random variable $U$ and can deduce the syntactic message $u$. In many applications of deep learning, the semantic features of input source data are extracted and mapped into a latent space which can be treated as the sample space of semantic variable.
\end{remark}

\section{Semantic Information Measures}
Based on synonymous mapping, we introduce the semantic information measures, such as, semantic entropy, up/down semantic mutual information, semantic channel capacity, and semantic rate distortion function.
\subsection{Semantic Entropy}
\begin{definition}
Given a discrete random variable $U$, the corresponding semantic variable $\tilde {U}$, and the synonymous mapping $f: \tilde{\mathcal{U}}\to\mathcal{U}$, the semantic entropy of semantic variable $\tilde {U}$ is defined by
\begin{equation}\label{semantic_entropy}
\begin{aligned}
&H_s(\tilde{U})=-\sum_{i_s=1}^{\tilde{N}}p\left(\mathcal{U}_{i_s}\right)\log p\left(\mathcal{U}_{i_s}\right) \\
&=-\sum_{i_s=1}^{\tilde{N}}\sum_{i\in\mathcal{N}_{i_s}}p\left(u_i\right) \log \left(\sum_{i\in\mathcal{N}_{i_s}}p\left(u_i\right)\right),
\end{aligned}
\end{equation}
\end{definition}
where $\mathcal{N}_{i_s}=\left\{N_{[1:(i_s-1)]}+1,\cdots,N_{[1:(i_s-1)]}+N_{i_s}\right\}$ is the index set associated with $\mathcal{U}_{i_s}$.
Essentially, semantic entropy is a functional of the distribution of $U$ and the synonymous mapping $f$. Similar to information entropy, semantic entropy also indicates the integrity attribute of the variable $\tilde{U}$ rather than single sample. Furthermore, it depends on the synonymous set partition determined by the mapping $f$. In Eq. (\ref{semantic_entropy}), the log is taken to the base 2 and we name the unit of semantic information as the semantic binary digit, that is, \textbf{sebit}.

\begin{lemma}\label{lemma2}
The semantic entropy is no more than the associated information entropy, that is,
\begin{equation}
H_s(\tilde{U})\leq H(U).
\end{equation}
\end{lemma}

We introduce the jointly/conditionally synonymous mapping as following.
\begin{definition}
Given a pair of discrete semantic variables $(\tilde {U},\tilde {V})$ and the corresponding random variable pairs $\left(U,V\right)$, $f_{u}: \tilde {\mathcal{U}} \to \mathcal{U}$ denotes the synonymous mapping from $\tilde {\mathcal{U}}$ to $\mathcal{U}$ and we have $f_{u}: \tilde{u}_{i_s} \to \mathcal{U}_{i_s}$ where $1\leq i_s\leq |\tilde {\mathcal{U}}|={\tilde{N}}_u$ and $1\leq i \leq\left| {\mathcal{U}}\right|=N_u$. Similarly, we can define the synonymous mapping $f_{v}: \tilde {\mathcal{V}} \to \mathcal{V}$. Furthermore, the jointly synonymous mapping $f_{uv}: \tilde {\mathcal{U}}\times \tilde{\mathcal{V}}\to \mathcal{U}\times{\mathcal{V}}$ is defined as
\begin{equation}
f_{uv}: (\tilde{u}_{i_s},\tilde{v}_{j_s}) \to \mathcal{U}_{i_s}\times \mathcal{V}_{j_s}.
\end{equation}
\end{definition}

The definition of semantic entropy can be further extended to a pair of semantic variables.
\begin{definition}
Given a pair of discrete semantic variables $(\tilde {U},\tilde {V})$ and the corresponding random variable pairs $\left(U,V\right)$ with a joint distribution $p(u,v)$, under the joint mapping $f_{uv}: \tilde {\mathcal{U}}\times \tilde{\mathcal{V}}\to \mathcal{U}\times{\mathcal{V}}$, the semantic joint entropy $H_s(\tilde {U},\tilde {V})$ is defined as
\begin{equation}\label{joint_semantic_entropy}
\begin{aligned}
H_s(\tilde {U},\tilde {V})=-\sum_{i_s=1}^{{\tilde{N}}_u}\sum_{j_s=1}^{{\tilde{N}}_v}&p\left(\mathcal{U}_{i_s}\times\mathcal{V}_{j_s}\right) \log p\left(\mathcal{U}_{i_s}\times\mathcal{V}_{j_s}\right)  \\
=-\sum_{i_s=1}^{{\tilde{N}}_u}\sum_{j_s=1}^{{\tilde{N}}_v} &\sum_{(u_i,v_j)\in \mathcal{U}_{i_s}\times \mathcal{V}_{j_s}} p\left(u_i,v_j\right)\\
\cdot \log &\sum_{(u_i,v_j)\in \mathcal{U}_{i_s}\times \mathcal{V}_{j_s}} p\left(u_i,v_j\right).
\end{aligned}
\end{equation}
\end{definition}

\begin{definition}
Given a pair of discrete semantic variables $(\tilde {U},\tilde {V})$ and the corresponding random variable pairs $\left(U,V\right)\sim p(u,v)$, under the conditional mapping $f_{v|u}: \tilde {\mathcal{V}}|u \to \mathcal{V}|u$, the semantic conditional entropy $H_s(\tilde {V}|U)$ is defined as
\begin{equation}
\begin{aligned}
H_s(\tilde {V}| U)&=-\sum_{i=1}^{N_u}\sum_{j_s=1}^{{\tilde{N}}_v}p\left(u_i\right)p\left(\mathcal{V}_{j_s}\left|u_i\right.\right) \log p\left(\mathcal{V}_{j_s}\left|u_i\right.\right)  \\
&=-\sum_{i=1}^{N_u}\sum_{j_s=1}^{{\tilde{N}}_v} \sum_{(u_i,v_j)} p\left(u_i,v_j\right) \log \sum_{v_j\in \mathcal{V}_{j_s}|u_i} p\left(v_j\left|u_i\right.\right).
\end{aligned}
\end{equation}
\end{definition}

\begin{lemma}\label{lemma3}
The semantic joint/conditional entropy is no more than the counterpart of CIT, that is,
\begin{equation}
\left\{
\begin{aligned}
H_s(\tilde{U},\tilde{V})&\leq H(U,V),\\
H_s(\tilde{U}|V)&\leq H(U|V).\\
\end{aligned}\right.
\end{equation}
\end{lemma}

The relationship between the semantic joint entropy and conditional entropy is indicated by the following theorem.
\begin{theorem}\label{theorem1}
(\textit{Chain Rule of Entropy}):
\begin{equation}\label{chain_rule}
\begin{aligned}
       H_s(\tilde{U})+H_s(\tilde{V}\left|U\right.) & \leq H_s(\tilde{U},\tilde{V})\\
\leq  H(U)+H_s(\tilde{V}\left|U\right.) & \leq H(U,V)
\end{aligned}
\end{equation}
\end{theorem}
\subsection{Semantic Mutual Information}

\begin{definition}\label{definition7}
Consider two semantic variables $\tilde{U}$ and $\tilde{V}$ and two associated random variables $U$ and $V$ with a joint probability mass function $p(u,v)$ and marginal probability mass function $p(u)$ and $p(v)$. Given the jointly synonymous mapping $f_{uv}: \tilde {\mathcal{U}}\times \tilde{\mathcal{V}}\to \mathcal{U}\times{\mathcal{V}}$, the up semantic mutual information $I^s(\tilde{U};\tilde{V})$ is defined as follows,
\begin{equation}
\begin{aligned}
I^s(\tilde{U};\tilde{V})=&-\sum_{i_s=1}^{{\tilde{N}}_u}\sum_{j_s=1}^{{\tilde{N}}_v}\sum_{(u_i,v_j)\in \mathcal{U}_{i_s}\times \mathcal{V}_{j_s}}p\left(u_i,v_j\right) \\
&\cdot \log \frac{p\left(u_i\right)p\left(v_j\right)}{\sum_{(u_i,v_j) \in \mathcal{U}_{i_s} \times \mathcal{V}_{j_s}}p\left(u_i,v_j\right)}\\
=&H(U)+H(V)-H_s(\tilde{U},\tilde{V}).
\end{aligned}
\end{equation}

Similarly, the down semantic mutual information $I_s(\tilde{U};\tilde{V})$ is defined as
\begin{equation}
\begin{aligned}
I_s (\tilde{U};\tilde{V})=&-\sum_{i_s=1}^{{\tilde{N}}_u}\sum_{j_s=1}^{{\tilde{N}}_v}\sum_{(u_i,v_j)\in \mathcal{U}_{i_s}\times \mathcal{V}_{j_s}}p\left(u_i,v_j\right)\\
 &\cdot \log \frac{\sum_{u_i \in \mathcal{U}_{i_s}}p\left(u_i\right) \sum_{v_j \in \mathcal{V}_{j_s}}p\left(v_j\right)}{p\left(u_i,v_j\right)}\\
=&H_s(\tilde{U})+H_s(\tilde{V})-H(U,V).
\end{aligned}
\end{equation}
\end{definition}
Note that the down semantic mutual information $I_s (\tilde{U};\tilde{V} )$ may be negative. Considering the practical case, we can set $(I_s (\tilde{U};\tilde{V} ))^{+}$.

We now show the relationship among these mutual information measures.
\begin{theorem}\label{theorem2}
(\textit{Chain Rule of Mutual Information}):
\begin{equation}
\begin{aligned}
      I_s (\tilde{U};\tilde{V}) &\leq H_s(\tilde{V})-H(V|U) \leq I (U;V )\\
                                           & \leq H (V )-H_s (\tilde{V}\left|U\right. ) \leq I^s (\tilde{U};\tilde{V} )\\
\end{aligned}
\end{equation}
\end{theorem}

\begin{table}[htbp]
\centering
\caption{Joint probability distribution of random variable pair $(U,V)$.} \label{JPDF_RV}
\scalebox{0.8}
{\begin{tabular}{|c|c|c|c|c|c|}
  \hline $(U,V)$     & $(u_1,v_1)$ &  $(u_1,v_2)$  & $(u_1,v_3)$ & $(u_1,v_4)$  & $(u_1,v_5)$  \\
  \hline $p(u,v)$     &   $0.05$       &       $0.1$       &    $0.15$      &       $0$         &       $0$         \\
  \hline $(U,V)$      & $(u_2,v_1)$ &  $(u_2,v_2)$  & $(u_2,v_3)$ & $(u_2,v_4)$  & $(u_2,v_5)$  \\
  \hline $p(u,v)$     &     $0.1$       &     $0.05$       &    $0.05$      &     $0.1$        &       $0$         \\
  \hline $(U,V)$     & $(u_3,v_1)$ &  $(u_3,v_2)$  & $(u_3,v_3)$ & $(u_3,v_4)$  & $(u_3,v_5)$  \\
  \hline $p(u,v)$     &     $0.1$       &     $0.05$       &        $0$       &       $0$         &      $0.05$     \\
  \hline $(U,V)$     & $(u_4,v_1)$ &  $(u_4,v_2)$  & $(u_4,v_3)$ & $(u_4,v_4)$  & $(u_4,v_5)$  \\
  \hline $p(u,v)$     &   $0.05$       &       $0$          &        $0$       &     $0.1$        &      $0.05$     \\
  \hline
\end{tabular}
}
\end{table}

\begin{table}[htbp]
\centering
\caption{Joint synonymous mapping of semantic variable pair $(\tilde{U},\tilde{V})$.} \label{JSmapping_SRV}
\scalebox{0.7}
{\begin{tabular}{|c|c|c|}
  \hline $f_{uv}$  & $({\tilde{u}}_1,{\tilde{v}}_1)\to\{(u_1,v_1),(u_2,v_1)\}$    & $({\tilde{u}}_1,{\tilde{v}}_2)\to\{(u_1,v_2),(u_2,v_2)\}$ \\
  \hline $p(\tilde{u},\tilde{v})$                   &       $0.15$            &            $0.15$        \\
  \hline $f_{uv}$  & $({\tilde{u}}_1,{\tilde{v}}_3)\to\{(u_1,v_3),(u_2,v_3)\}$ & $({\tilde{u}}_1,{\tilde{v}}_4)\to\{(u_1,v_4),(u_1,v_5),(u_2,v_4),(u_2,v_5)\}$ \\
  \hline $p(\tilde{u},\tilde{v})$                   &      $0.2$         &         $0.1$  \\
  \hline $f_{uv}$  & $({\tilde{u}}_2,{\tilde{v}}_1)\to\{(u_3,v_1),(u_4,v_1)\}$ & $(\tilde{u}_2,\tilde{v}_2)\to\{(u_3,v_2),(u_4,v_2)\}$ \\
  \hline $p(\tilde{u},\tilde{v})$                   &  $0.15$    &  $0.05$ \\
  \hline $f_{uv}$  & $(\tilde{u}_2,\tilde{v}_3)\to\{(u_3,v_3),(u_4,v_3)\}$ & $(\tilde{u}_2,\tilde{v}_4)\to\{(u_3,v_4),(u_4,v_4),(u_3,v_5),(u_4,v_5)\}$ \\
   \hline $p(\tilde{u},\tilde{v})$                   &   $0$   &  $0.2$  \\
  \hline
\end{tabular}
}
\end{table}

\begin{table}[htbp]
\centering
\caption{Conditional synonymous mapping of semantic variable $\tilde{U}|V$.} \label{CSmapping_SRV}
\scalebox{0.7}
{\begin{tabular}{|c|c|c|c|}
  \hline $\tilde{U}|V$     & $\tilde{u}_1|v_1\to \{u_1,u_2\}|v_1 $ &  $\tilde{u}_1|v_2\to \{u_1,u_2\}|v_2$ & $\tilde{u}_1|v_3\to \{u_1,u_2\}|v_3$ \\
  \hline $p(\tilde{u}|v)$         &   $0.5$       &       $0.75$       &    $1$     \\
  \hline $\tilde{U}|V$     & $\tilde{u}_1|v_4\to \{u_1,u_2\}|v_4$ & $\tilde{u}_1|v_5\to \{u_1,u_2\}|v_5$ &    \text{ } \\
  \hline $p(\tilde{u}|v)$         &       $0.5$         &       $0$      &   \text{ }   \\
  \hline $\tilde{U}|V$     & $\tilde{u}_2|v_1\to \{u_3,u_4\}|v_1 $ &  $\tilde{u}_2|v_2\to \{u_3,u_4\}|v_2$ & $\tilde{u}_2|v_3\to \{u_3,u_4\}|v_3$    \\
  \hline $p(\tilde{u}|v)$         &     $0.5$       &     $0.25$       &        $0$    \\
  \hline $\tilde{U}|V$     & $\tilde{u}_2|v_4\to \{u_3,u_4\}|v_4$ & $\tilde{u}_2|v_5\to \{u_3,u_4\}|v_5$  &    \text{ } \\
  \hline $p(\tilde{u}|v)$         &       $0.5$         &      $1$  &    \text{ }   \\
  \hline
\end{tabular}
}
\end{table}

\begin{table}[htbp]
\centering
\caption{Conditional synonymous mapping of semantic variable $\tilde{V}|U$.} \label{CSmapping_SRV2}
\scalebox{0.65}
{\begin{tabular}{|c|c|c|c|c|}
  \hline $\tilde{V}|U$     & $\tilde{v}_1|u_1\to \{v_1\}|u_1 $ &  $\tilde{v}_1|u_2\to \{v_1\}|u_2$ & $\tilde{v}_1|u_3\to \{v_1\}|u_3$ & $\tilde{v}_1|u_4\to \{v_1\}|u_4$  \\
  \hline $p(\tilde{v}|u)$         &   $1/6$       &       $1/3$       &    $0.5$      &       $0.25$           \\
  \hline $\tilde{V}|U$     & $\tilde{v}_2|u_1\to \{v_2\}|u_1 $ &  $\tilde{v}_2|u_2\to \{v_2\}|u_2$ & $\tilde{v}_2|u_3\to \{v_2\}|u_3$ & $\tilde{v}_2|u_4\to \{v_2\}|u_4$  \\
  \hline $p(\tilde{v}|u)$         &   $1/3$       &       $1/6$       &    $0.25$      &       $0$             \\
  \hline $\tilde{V}|U$     & $\tilde{v}_3|u_1\to \{v_3\}|u_1 $ &  $\tilde{v}_3|u_2\to \{v_3\}|u_2$ & $\tilde{v}_3|u_3\to \{v_3\}|u_3$ & $\tilde{v}_3|u_4\to \{v_3\}|u_4$  \\
  \hline $p(\tilde{v}|u)$         &   $0.5$       &       $1/6$       &    $0$           &       $0$             \\
  \hline $\tilde{V}|U$     & $\tilde{v}_4|u_1\to \{v_4,v_5\}|u_1 $ &  $\tilde{v}_4|u_2\to \{v_4,v_5\}|u_2$ & $\tilde{v}_4|u_3\to \{v_4,v_5\}|u_3$ & $\tilde{v}_4|u_4\to \{v_4,v_5\}|u_4$  \\
  \hline $p(\tilde{v}|u)$         &   $0$         &       $1/3$        &    $0.25$      &       $0.75$         \\
  \hline
\end{tabular}
}
\end{table}

\begin{table}[htbp]
\centering
\caption{Synonymous mappings of semantic variables $\tilde{U}$ and $\tilde{V}$.} \label{Smapping_SRV}
\scalebox{0.65}
{
\begin{tabular}{|c|c|c|}
  \hline $f_{u}$               & $\tilde{u}_1\to\{u_1,u_2\}$  &  $\tilde{u}_2\to\{u_3,u_4\}$  \\
  \hline $p(\tilde{u})$           &           $0.6$                        &            $0.4$         \\
  \hline
\end{tabular}
\begin{tabular}{|c|c|c|c|c|}
  \hline $f_{v}$         & $\tilde{v}_1 \to\{v_1\}$  & $\tilde{v}_2 \to\{v_2\}$ & $\tilde{v}_3 \to\{v_3\}$ & $\tilde{v}_4 \to\{v_4,v_5\}$ \\
  \hline $p(\tilde{v})$    &               $0.3$               &            $0.2$                   &                $0.2$              &               $0.3$ \\
  \hline
\end{tabular}
}
\end{table}

\begin{example}
Table \ref{JPDF_RV} gives a joint probability distribution of random variable pair $(U,V)$. Table \ref{JSmapping_SRV} illustrates the distribution of the associated semantic variable pair $(\tilde{U},\tilde{V})$ under a joint synonymous mapping $f_{uv}$. Table \ref{CSmapping_SRV} and Table \ref{CSmapping_SRV2} give the conditional distribution of the semantic variables $\tilde{U}|V$ and $\tilde{V}|U$. The marginal distributions of the semantic variable $\tilde{U}$ and $\tilde{V}$ are depicted in Table \ref{Smapping_SRV}.

The joint entropy of $(U,V)$ is calculated as $H(U,V)=3.5842\text{ bits}$. The semantic joint entropy of $(\tilde{U},\tilde{V})$ is $H_s(\tilde{U},\tilde{V})=$ $-\sum_{i_s=1}^{2}\sum_{j_s=1}^{4} p(\tilde{u}_{i_s},\tilde{v}_{j_s})\log p(\tilde{u}_{i_s},\tilde{v}_{j_s})$ $=2.7087 \text{ sebits}$.

The conditional entropies are $H(U|V)=1.3377 \text{ bits}$ and $H(V|U)=1.6132 \text{ bits}$. Correspondingly, the semantic conditional entropies are $H_s(\tilde{U}|V)=-\sum_{i_s=1}^{2}\sum_{j=1}^{5} p(\tilde{u}_{i_s},v_j)$\\$\cdot\log p(\tilde{u}_{i_s}|v_j)$ $=0.6623 \text{ sebits}$ and $H_s(\tilde{V}|U)$ $=-\sum_{j_s=1}^{4}\sum_{i=1}^{4} p(\tilde{v}_{j_s},u_i)\log p(\tilde{v}_{j_s}|u_i)$ $=1.4755 \text{ sebits}$. Thus we have $H_s(\tilde{U}|V)=0.6623 \text{ sebits}<H(U|V)=1.3377 \text{ bits}$ and $H_s(\tilde{V}|U)=1.4755 \text{ sebits}<H(V|U)=1.6132 \text{ bits}$.

The entropies of random variables $U$ and $V$ are calculated as $H(U)=H(0.3,0.3,0.2,0.2)=1.971 \text{ bits}$ and $H(V)=H(0.3,0.2,0.2,0.2,0.1)=2.2464 \text{ bits}$ respectively. Then, the semantic entropies of $\tilde{U}$ and $\tilde{V}$ are $H_s(\tilde{U})=H_s(0.6,0.4)=0.971 \text{ sebits}$ and $H_s(\tilde{V})=H_s(0.3,0.2,0.2,0.3)=1.971 \text{ sebits}$ respectively.

According to the joint distribution in Table \ref{JPDF_RV}, we calculate the mutual information between $U$ and $V$ as $I(U;V)=H(U)+H(V)-H(U,V)=0.6332 \text{ bits}$. Correspondingly, by using the joint distribution in Table \ref{JSmapping_SRV}, we can compute the up semantic mutual information as $I^s(\tilde{U};\tilde{V})=H(U)+H(V)-H_s(\tilde{U};\tilde{V})=1.5087 \text{ sebits}$.

Similarly, by using the distribution in Table \ref{Smapping_SRV}, we compute the down semantic mutual information as $I_s(\tilde{U};\tilde{V})=H_s(U)+H_s(V)-H(U;V)=-0.6422 \text{ sebits}$. Consider the non-negative requirement, we set $I_s(\tilde{U};\tilde{V})=\max\{-0.6422,0\}=0 \text{ sebits}$. By using the distribution in Table \ref{CSmapping_SRV} and Table \ref{CSmapping_SRV2}, we calculate that $H(V)-H_s(\tilde{V}\left|U\right.)=0.7709 \text{ sebits}$ and $H(U)-H_s(\tilde{U}\left|V\right.)=1.3087 \text{ sebits}$. So we conclude that $I_s(\tilde{U};\tilde{V})=0 \text{ sebits}=H_s(\tilde{U})-H(U\left|V\right.)= \max\{-0.3667,0\}=0 \text{ sebits}<H_s(\tilde{V})-H(V\left|U\right.)=0.3578 \text{ sebits}<I(U;V)=0.6332 \text{ bits}<H(V)-H_s(\tilde{V}\left|U\right.)=0.7709 \text{ sebits}<H(U)-H_s(\tilde{U}\left|V\right.)=1.3087 \text{ sebits}<I^s(\tilde{U};\tilde{V})=1.5087 \text{ sebits}$.
\end{example}
\subsection{Semantic Channel Capacity and Semantic Rate Distortion}
\begin{definition}
Given a discrete memoryless channel $\left\{\tilde{\mathcal{X}},\mathcal{X},\mathcal{Y},\tilde{\mathcal{Y}},f_{xy}, p(Y|X)\right\}$, the semantic channel capacity is defined as
\begin{equation}
\begin{aligned}
C_s=&\max_{p(x)} I^s(\tilde{X};\tilde{Y})=\max_{p(x)}\left[H(X)+H(Y)-H_s(\tilde{X},\tilde{Y})\right]
\end{aligned}
\end{equation}
where the maximum is taken over all possible input distribution $p(x)$.
\end{definition}

By Theorem \ref{theorem2}, the channel mutual information is no more than the up semantic mutual information, that is, $I(X;Y)\leq I^s(\tilde{X};\tilde{Y})$. Thus, we conclude that the channel capacity is no more than the semantic capacity, i.e., $C\leq C_s$.

\begin{definition}
Given an i.i.d. source $X$ with distribution $p(x)$, the associated semantic source $\tilde{X}$, and the semantic distortion function $d_s(\tilde{x}_{i_s},\hat{\tilde{x}}_{j_s})$, the semantic rate distortion is defined as,
\begin{equation}
\begin{aligned}
R_s(D)=&\min_{p(\hat{x}\left|x\right.)\in P_D}I_s(\tilde{X};\hat{\tilde{X}})\\
=&\min_{p(\hat{x}\left|x\right.)\in P_D} \left[H_s(\tilde{X})+H_s(\hat{\tilde{X}})-H(X,\hat{X})\right],
\end{aligned}
\end{equation}
where $P_D=\left\{p(\hat{x}\left|x\right.): \bar{d}_s=\mathbb{E}\left[d_s\left(\tilde{x},\hat{\tilde{x}}\right)\right] \leq D\right\}$ denotes the test channel set.
\end{definition}

\section{Semantic Coding Theorems}

\subsection{Asymptotic Equipartition Property and Synonymous Typical Set}
Similar to the definition of syntactically typical set $A_{\epsilon}^{(n)}$ in \cite{Book_Cover}, we can define the semantically typical set $\tilde{A}_{\epsilon}^{(n)}$. Furthermore, we define the synonymous typical set $B_{\epsilon}^{(n)}\left(\tilde{u}^n\right)$ as following.

\begin{definition}
Given a specific typical sequence $\left\{\tilde{u}^n\right\}$, the synonymous typical set $B_{\epsilon}^{(n)}\left(\tilde{u}^n\right)$ with the syntactically typical sequences $\left\{u^n\right\}$ is defined as the set of $n$-sequences with the difference of empirical entropies  $\epsilon$-close to the difference of true entropies, that is,
\begin{equation}
\begin{aligned}
B_{\epsilon}^{(n)}\left(\tilde{u}^n\right)=\bigg\{ u^n\in \mathcal{U}^n: &\left|-\frac{1}{n}\log p\left(u^n\right)-H(U)\right|<\epsilon,\\
&\left|-\frac{1}{n}\log p\left(\tilde{u}^n\right)-H_s(\tilde{U})\right|<\epsilon,\\
&\left|-\frac{1}{n}\log p(\tilde{u}^n\to u^n)\right.\\
&\left.\left.-\left(H(U)-H_s(\tilde{U})\right)\right|<\epsilon\right\},
\end{aligned}
\end{equation}
where $p(u^n)\!\!=\!\!\prod_{k=1}^{n} p(u_k)$ and $ p(\tilde{u}^n)\!\!=\!\!\prod_{k=1}^{n} \sum_{u_k\in \mathcal{U}_{\tilde{u}_k}}\!\!\!\!p(u_k)$. Here the probability $p(\tilde{u}^n\to u^n)$ is defined as
\begin{equation}
p(\tilde{u}^n\to u^n)=\left\{
\begin{aligned}
&\frac{p\left(u^n\right)}{p\left(\tilde{u}^n\right)},&\text{ if } u^n=f^n(\tilde{u}^n),\\
&0, &\text{otherwise}.
\end{aligned}\right.
\end{equation}
\end{definition}

Essentially, $B_{\epsilon}^{(n)}\left(\tilde{u}^n\right)$ is an equivalence class of the synonymous typical sequences. For the syntactically typical set $A_{\epsilon}^{(n)}$, we can obtain a quotient set $A_{\epsilon}^{(n)}/f^n=\left\{B_{\epsilon}^{(n)}(\tilde{u}^n)\right\}$ and construct an one-to-one mapping between $\tilde{A}_{\epsilon}^{(n)}$ and $A_{\epsilon}^{(n)}/f^n$. In \cite{SemInfo_theory}, we prove that $ 2^{n\left(H(U)-H_s(\tilde{U})-\epsilon\right)}\leq\left|B_{\epsilon}^{(n)}\left(\tilde{u}^n\right)\right| \leq 2^{n\left(H(U)-H_s(\tilde{U})+\epsilon\right)}$ for sufficiently large $n$.

\subsection{Semantic Source Coding Theorem}
\begin{theorem}
(Semantic Source Coding Theorem)

Given the semantic source $\tilde{U}$ and the syntactic source $U$ with the synonymous mapping $f:\tilde{\mathcal{U}}\to \mathcal{U}$, for each code rate $R>H_s(\tilde{U})$, there exists a series of $\left(2^{n(R+R_s)},n\right)$ codes, when code length $n$ tends to sufficiently large, the error probability is close to zero, i.e. $P_e^{(n)}\to 0$. On the contrary, if $R<H_s(\tilde{U})$, then for any $\left(2^{n(R+R_s)},n\right)$ code, the error probability tends to $1$ with $n$ sufficiently large.
\end{theorem}
\begin{IEEEproof}
First, we prove the direct part of the theorem. We select $\epsilon>0$ and construct a one-to-one mapping from $\tilde{A}_{\epsilon}^{(n)}$ to $\mathcal{C}/f^n$. So for sufficiently large $n$, we have $\left(1-\epsilon\right) 2^{n\left(H_s(\tilde{U})-\epsilon\right)}\leq\left|\mathcal{C}/f^n\right|=2^{nR}=\left|\tilde{A}_{\epsilon}^{(n)}\right| \leq 2^{n\left(H_s(\tilde{U})+\epsilon\right)}$.
Therefore, the semantic code rate satisfies $\frac{1}{n}\log\left(1-\epsilon\right)+H_s(\tilde{U})-\epsilon \leq R \leq H_s(\tilde{U})+\epsilon$.
Also by semantic AEP, it follows that $P_e^{(n)}=\text{Pr}(\tilde{U}^n\notin \tilde{A}_{\epsilon}^{(n)})<\epsilon$.

By synonymous typicality, the size of synonymous set satisfies $1\leq\left|\mathcal{C}_s\right|=2^{nR_s}\leq \left|B_{\epsilon}^{(n)}\left(\tilde{u}^n\right)\right| \leq 2^{n\left(H(U)-H_s(\tilde{U})+\epsilon\right)}$.
Therefore the syntactic code rate satisfies
\begin{equation}\label{equation74}
\frac{1}{n}\log\left(1-\epsilon\right)+R_s+H_s(\tilde{U})-\epsilon \leq R' \leq H_s(\tilde{U})+R_s+\epsilon .
\end{equation}

When the synonymous set $\mathcal{C}_s$ only has one codeword, that is, $R_s=0$, letting $\epsilon\to 0$, both the code rate $R$ and $R'$ tend to $H_s(\tilde{U})$, while $P_e^{(n)}$ tends to $0$. On the other hand, substituting $R_s=H(U)-H_s(\tilde{U})$ into (\ref{equation74}) and letting $\epsilon\to 0$, $R\to H_s(\tilde{U})$ and $R'\to H(U)$. So we prove the direct part of theorem. The converse part is proved in \cite{SemInfo_theory}.
\end{IEEEproof}

\subsection{Semantic Channel Coding Theorem}
We now give the formal description of semantic channel coding theorem.
\begin{theorem}\label{Sem_CCT}
(Semantic Channel Coding Theorem)

Given the semantic channel $\left\{\tilde{\mathcal{X}},\mathcal{X},\mathcal{Y},\tilde{\mathcal{Y}},f_{xy}, p(Y|X)\right\}$, for each code rate $R<C_s$, there exists a sequence of $\left(2^{n(R+R_s)},n\right)$ codes consisting of synonymous codeword set with the rate $0\leq R_s \leq H(X,Y)-H_s(\tilde{X},\tilde{Y})$, when code length $n$ tends to sufficiently large, the error probability tends to zero, i.e. $P_e^{(n)}\to 0$.

On the contrary, if $R>C_s$, then for any $\left(2^{n{(R+R_s)}},n\right)$ code, the error probability tends to $1$ with sufficiently large $n$.
\end{theorem}

In \cite{SemInfo_theory}, we introduce the jointly synonymous typical set $B_{\epsilon}^{(n)}\left(\tilde{x}^n,\tilde{y}^n\right)$ and use the random coding and jointly typical decoding to prove this theorem. We can give an intuitive idea to explain the proof. For each syntactically typical input $n$-sequence $X^n$, there are approximately $2^{nH_s(\tilde{Y}|X)}$ possible output sequences $Y^n$. Since the total number of syntactically typical sequences $Y^n$ is about $2^{nH(Y)}$, this typical set can be divided into subsets with size $2^{nH_s(\tilde{Y}|X)}$ corresponding to the distinct input sequence $X^n$. So the total number of disjoint sets is about $2^{n(H(Y)-H_s(\tilde{Y}|X))}$. If we further consider the synonymous mapping of input sequences, due to $H(Y)-H_s(\tilde{Y}|X)\leq I^s(\tilde{X};\tilde{Y})$ by Theorem \ref{theorem2}, we can send at most $2^{nI^s(\tilde{X};\tilde{Y})}$ distinguishable sequences of length $n$.

\subsection{Semantic Rate-Distortion Coding Theorem}
We now give the formal description of semantic rate-distortion coding theorem.
\begin{theorem}\label{Sem_RDT}
(Semantic Rate-Distortion Coding Theorem):

Given an i.i.d. syntactic source $X\sim p(x)$ with the associated semantic source $\tilde{X}$ under the synonymous mapping $f$ and the bounded semantic distortion function $d_s(\tilde{x},\hat{\tilde{x}})$, for each code rate $R>R_s(D)$, there exists a sequence of $\left(2^{n(R+R_s)},n\right)$ codes, when code length $n$ tends to sufficiently large, the semantic distortion satisfies $\mathbb{E}d_s(\tilde{X},\hat{\tilde{X}})<D$.

On the contrary, if $R<R_s(D)$, then for any $\left(2^{n{(R+R_s)}},n\right)$ code, the semantic distortion meets $\mathbb{E}d_s(\tilde{X},\hat{\tilde{X}})>D$ with sufficiently large $n$.
\end{theorem}

This theorem is proved by using jointly typical encoding in \cite{SemInfo_theory}. We also give an intuitive idea to explain the proof. The total number of semantically typical sequences $\hat{\tilde{X}}^n$ is about $2^{nH_s(\hat{\tilde{X}})}$. For each syntactically typical sequence $X^n$, there are approximately $2^{nH(\hat{X}|X)}$ possible representative sequences $\hat{X}^n$. Therefore the total number of disjoint representative sets is about $2^{n(H_s(\hat{\tilde{X}})-H(\hat{X}|X))}$. If we further consider the synonymous mapping of source sequences, due to $I_s(\tilde{X};\hat{\tilde{X}})\leq H_s(\hat{\tilde{X}})-H(\hat{X}|X)$ by Theorem \ref{theorem2}, we can use at least $2^{nI_s(\tilde{X};\hat{\tilde{X}})}$ distinguishable sequences to represent the semantic source under the distortion contraint.

\section{Conclusion}
In this paper, we develop an information-theoretic framework of semantic communication. We start from the synonym, a fundamental property of semantic information, to build the semantic information measures. Then we extend the asymptotic equipartition property to the semantic sense and introduce the synonymous typical set to prove the semantic source coding theorem, semantic channel coding theorem, and semantic rate distortion coding theorem. All these works uncover the critical features of semantic communication and constitute the theoretic basis of semantic information theory.

\section*{Acknowledgement}
This work is supported by the National Natural Science Foundation of China (No. 62293481, 62071058).
\IEEEtriggeratref{4}


\end{document}